# REPUTATION-BASED ATTACK-RESISTANT COOPERATION STIMULATION (RACS) FOR MOBILE AD HOC NETWORKS

Anuradha Banerjee[1] and Paramartha Dutta[2]

[1]Department of Computer Applications, Kalyani Govt. Engg. College, Kalyani, West Bengal, India
anuradha79bn@gmail.com

[2]Department of Computer and System Science, Visva-Bharati University, Santiniketan, India
paramartha.dutta@gmail.com

## ABSTRACT

*In mobile ad hoc networks (MANET), nodes usually belong to different authorities and pursue different goals. In order to maximize their own performance, nodes in such networks tend to be selfish and are not willing to forward packets for benefit of others. Meanwhile, some nodes may behave maliciously and try to disrupt the network through wasting other nodes resources in a very large scale. In this article, we present a reputation-based attack resistant cooperation stimulation (RACS) system which ensures that damage caused by malicious nodes can be bounded and cooperation among the selfish nodes can be enforced. Mathematical analyses of the system as well as the simulation results have confirmed effectiveness of our proposed system. RACS is completely self-organizing and distributed. It does not require any tamper-proof hardware or central management policy.*

## 1. INTRODUCTION

An ad hoc network is a group of mobile nodes communicating with each other in single/multiple hops, without any centralized administration or fixed backbone network infrastructure [1-10]. Along with emergency and military situations, ad hoc networks are being envisaged in civilian usage also. On account of these diverse categories of applications, nodes usually belong to different authorities and may not pursue a common goal. Moreover, in order to save limited resources such as battery power, nodes may tend to be selfish and disagree to unconditionally forward packets of others. One possible way to stimulate cooperation is to use payment-based methods. An approach was proposed by using virtual currency called 'nuglets' as payments for packet forwarding, which was then improved in using credit counters [1]. However, tamper-proof hardware is required at each node to count the credits. Sprite [2] was proposed to encourage cooperation among network elements or nodes. Although, it releases the requirement of a tamper-proof hardware, a centralized credit-clearance service trusted by all nodes, is recognized as indispensable. This violates the desirable fully distributed structure of ad hoc network.

The first reputation based system for ad hoc networks was proposed to mitigate a node's misbehaviour, where each network element launches a watchdog to monitor its neighbours packet forwarding activities. After that, CORE [3] and CONFIDANT [4] were proposed to detect and isolate misbehaving nodes and thus, make it unattractive to deny cooperation. These schemes treat all misbehaving nodes in the same manner, irrespective of their degree of selfish or malicious activity. Sometimes, innocent cooperating nodes have to suffer due to misbehaviour of other nodes and incapability of the underlying mechanism to identify the culprit(s). These experiences clearly show that, before ad hoc networks are successfully deployed, security concerns must be addressed in full proof manner. Most of the schemes proposed in the literature to address security issues in ad hoc networks, focus on preventing attackers from entering the network through secure key distribution and neighbour discovery. These are unable to handle the situation where malicious nodes successfully enter the network. The issue is quite worthy of attention and technically challenging since access control in autonomous ad hoc networks is usually very loose.

George Karakostas et. Al. proposed a protocol for emergency connectivity [5] in ad hoc networks with selfish nodes. This is inspired by the CONFIDANT [4] protocol where reputation mechanism is implemented through the ability of any node to define a threshold of tolerance for any of its neighbours and to cut the connection to any of these neighbours that refuse to forward an amount of flow above the threshold. The protocol is modelled as a game and necessary and sufficient conditions are given for the existence of connected Nash equilibria. However, it remains silent about the existence of malicious nodes and cannot differentiate between the facts when a node intentionally drops a packet or drops a packet due to scarcity of battery power.

The technique of activity-based overhearing [6] is proposed by Frank Kargl in [6]. Each router tries to overhear forwarding of data packets by its next hop. The date of last regular activity of a downlink neighbour is stored in a table. When it sends a packet to another node and cannot detect forwarding of the packet by the next relaying node in the path, this is esteemed a selfish behaviour only when there has been a recent regular activity by this node. The monitoring node triggers an alarm only when it detects a certain number of packets being dropped within a certain timeframe.

In the present article, we have separately considered the various attacking mechanisms used by selfish and malicious nodes. Objective of a selfish node is to maximize the benefits it can extract out of the network while preserving its resources as much as possible. On the other hand, objective of a malicious node is to maximize the damage it can cause to the network (may be without any explicit advantage), at the cost of its own resources, till the expense is tolerable. Possible types of attacking procedures have been rigorously analyzed and sufficient intelligence is artificially embedded within the nodes to prevent such harmful activities. Cooperation is thereby enforced as the best policy of a node for survival in a mobile ad hoc network. Moreover, the system of RACS is completely distributed and does not require any tamper-proof hardware. It can identify the malfunctioning nodes with high percentage of accuracy and determine their punishment according to their degree of un-cooperating attitude, both during transmission of data packets and route discovery.

This paper is organized as follows. Various attacking schemes employed in ad hoc networks are mentioned in section 2. Section 3 describes the system model in detail and analyzes behaviour of associated nodes under different circumstances. Communication algorithm of RACS is provided in section 4. Simulation studies appear in section 5 and finally, section 6 concludes the paper.

## 2. SELFISH AND MALICIOUS ACTIVITIES

In an autonomous ad hoc network, each node is equipped with a battery of limited power supply and may act as a service provider. To extend own life, a node may exhibit selfish behaviour by benefiting from resources provided by other nodes, without, in return, making available resources of their own devices. RACS encourages selfish nodes (directly or as witnesses) in forwarding activities by rewarding them with significant increment in reputation. Malicious fellows, on the other hand, do not care about the rewards and cause harm to the network in multi-faced manners as much as possible, till the expense is tolerable. If a node is found to be malfunctioning deliberately, RACS recommends it to be isolated from the network. In general, attacking mechanisms in ad hoc networks are mentioned below:

- Emulating Link Breakage – When a node $n_a$ requires to transmit a message packet to the next router $n_b$ on a communication path, $n_b$ may simply keep silent to $n_a$ to let it believe that its successor in the route has gone out of its radio-range. This phenomenon is termed as link breakage attack. Details about this attacking mechanism and its treatment in RACS appear in section 3.2.
- Deliberately Delaying The Traffic – A node may be idle in forwarding of messages and deliberately introduce delay. Detection of this kind of attack and corresponding adjustment in reputation of the guilty nodes, are demonstrated in section 3.3.
- Injecting Enormous Traffic – Malicious nodes can inject an overwhelming amount of traffic (frequent flooding of data or route-request packets) to overload the network and consume valuable energy of others. This is elaborated in section 3.4.
- Collusion Attack – Attackers may work together to improve their attacking ability. Section 3.5 is devoted for its treatment in RACS.
- Slander Attack – Attackers can also recommend undue blacklisting (minimizing of reputation) of others. Discussion about it is elaborated in section 3.5.
- Masking Attack – A node may pretend to be another one while transmitting messages. Details about it are provided in section 3.6.

Please note that, RACS is a reputation-based system. Whenever message forwarding requests come to a node, it prioritizes them according to reputation of their respective sources and processes in that order. If more than one source has same reputation, they are arranged in first-come-first-serve basis in the message queue. Selfish nodes need not be punished because, as soon as acknowledgement arrives from destination of a communication path to the associated source node, RACS increments reputation of all routers to the source. Malicious nodes are subjected to detection and punishment (even blacklisting i.e. minimization of reputation almost throughout the network). RACS stimulates malicious nodes to detect malfunctioning of other malicious fellows instead of conspiring against selfish ones.

## 3. SYSTEM MODEL AND ANALYSIS

Different nodes are assumed to have different transmission power and be equipped with two antennas – i) one omni-directional antenna essential for broadcasting and ii) one uni-directional antenna for unicasting messages to the nodes with very high reputation (also termed as friends) to the predecessor node (router/source). Omni-directional antenna is used to unicast messages to those who are not known to be friends of the transmitting node. Each node has a unique link-level address which cannot be changed and is included in all messages that it transmits / forwards. RACS ensures that current timestamp is a part of every message generated in an ad hoc network. A fuzzy controller named Adaptive Penalty Decider (APD) is embedded in every node to incorporate rationality in decision-making as far as penalizing others is concerned. Unless on the verge of complete exhaustion, each node $n_i$ transmits HELLO messages at a regular interval $\tau_{hello}(i)$ (this interval is non-uniform). Receiving a HELLO message, all downlink neighbours $n_j$ of $n_i$ send acknowledgement (ACK) to $n_i$ which contains information about identification number, present coordinate position, radio-range, current timestamp, hello interval and minimum time required to process and forward a message, of ownself and downlink neighbours (conforming to last HELLO message transmitted by $n_j$). All these attributes are automatically included in all data and route-request packets generated by a node. Values of these attributes of $n_i$, along with the same of its most recent set of uplink and downlink neighbours are included in HELLO message of $n_i$. After transmitting or relaying a message, a node waits for a maximum interval of $\tau'$ seconds to receive acknowledgement. If no ACK reply arrives within the mentioned time-span, the message is resent. This may be repeated at most 3 times (one initial transmission, two retransmissions). Current timestamp is embedded in all messages transmitted in the network. Each node is equipped with a message queue. Size of message queue of $n_i$ is denoted as $m_i$. Received messages are stored here for processing and forwarding, which takes place in order of reputation of the associated nodes to $n_i$. Whenever malicious behaviour of a node is detected with proof, the node is blacklisted to the detector as well as to all those nodes who receive the allegation intimation along with supportive digital documents broadcasted by the detector. If a node is blacklisted, then all messages generated by it would be discarded without consideration, by all other nodes in the network. RACS also assumes that no node is capable of distorting content of a message transmitted by any other node.

Assuming that $\Phi$ is the present set of all nodes in the network, maximum and minimum reputation of a node are $|\Phi|$ and $-|\Phi|$, respectively. If reputation of a node $n_j$ to another node $n_i$ becomes less than $-|\Phi|$, $n_j$ is blacklisted to $n_i$ although the information cannot be broadcasted throughout the network since explicit proof of non-cooperation is not available (complains without document are treated as slander attack and nodes raising slander attack are blacklisted to all those who receive allegation from it). In this situation, messages generated by $n_j$ are treated by $n_i$ with least priority, $n_i$ does not assign any message forwarding task to $n_j$. On the other hand, if a malicious node $n_j$ is caught red-handed, it is blacklisted network-wide and no node entertains its message forwarding requests and all complains generated by it are completely ignored.

### 3.1. The forwarding procedure

Characteristics of ad hoc network entities or nodes are mentioned below:

- Current set of neighbours of a node $n_i$ is denoted as $N(i)$. Although this is a function of time, we have ignored the fact in our notation for simplicity of representation.

- Class of network elements may be broadly classified into i) selfish nodes and ii) malicious nodes. A selfish node is either a) supportive or b) interrupt-driven. Objective of a selfish node is to save its own resources and maximize its reputation to others, as much as possible. A supportive node is one who renders service to others without explicit request or responsibility. For each such enthusiastic cooperation, reputation of a supportive node is set to $\alpha|\Phi|$ ($\alpha > 1$) to all nodes who are benefited from it. For example, consider a communication path $n_i \rightarrow n_j \rightarrow n_k$ which satisfies the following conditions:

  a) $n_i \notin N(j)$
  b) There exists a node $n_l$ such that $n_j \in N(l)$, $n_l \in N(j)$, $n_i \in N(l)$ and $n_l \in N(i)$.

  Then, $n_l$ is termed as a witness in the path from $n_i$ to $n_j$. After $n_j$ forwards a message sent by predecessor router $n_i$, $n_l$ also receives a copy of that which it bicasts to both $n_i$ and $n_j$. As $n_i$ receives the broadcasted packet, it compares the message sent by it and received copy of $n_l$. If they are identical, then honest and responsible service of $n_j$ is confirmed by unforeseen contribution of $n_l$ and thereby, requirement of an explicit multi-hop acknowledgement initiated by $n_j$, is eliminated. Here $n_l$ is an example of supportive node and its response is treated as implicit acknowledgement. If a selfish node is not supportive, then it is interrupt-driven. On the other hand, objective of a malicious node is to consume network resources and conspire against others as much as possible, till the expense is bearable.

- In a communication path $n_i \rightarrow n_j \rightarrow n_k$, if $n_i \notin N(j)$ then $n_j$ bicasts its packet to be forwarded before retransmissions begin with its two destinations being itself and $n_k$. $n_j$

deliberately transmits its received copy to $n_i$ provided it is about to leave the radio-circle of $n_i$. If a proper ACK arrives from destination to $n_i$ where $n_i$ is either a source or a router, reputation of relaying nodes who are successors of $n_i$, get incremented by $\alpha$ to their respective predecessors. Reputation of generator of the message reduces by $\alpha^2$ to every router. Otherwise the situation is identified as link breakage and treated in section 3.2. Here it may please be noted that ACK messages are also bicasted with ownself being the additional receiver. If a node is on the verge of going outside the radio-circle of its predecessor and properly informs the previous router about it, RACS ensures that its predecessor does not complain to other network elements against it.

- A node never complains that its successor has not responded when the successor has already acknowledged. This is true for malicious nodes also.

- A node does not complain against its successor provided the received copy of bicasted transmission has already arrived at it. This is true for malicious nodes also.

- The routers for whom neither predecessor is within its radio-range nor witness support is available, append received copies of message bicasted by itself, along with acknowledgement during its traversal from destination to source. This copy is extracted by the respective predecessor and replaced by its own received copy of bicasted message. If no acknowledgement arrives from destination after a pre-computed time interval, each node voluntarily returns received copy of its bicasted message to the predecessor.

## 3.2. Link Breakage Attack

Consider a communication path … $n_i \rightarrow n_j \rightarrow n_k$ …Assume that $n_i$ sends a data packet at time t1 to $n_j$ and did not receive any acknowledgement till $t1+3\tau'$ i.e. after two retransmissions. Immediately $n_i$ issues a route-request message to $n_k$ excluding $n_j$. If any such alternative path is found, $n_i$ inquires $n_k$ about HELLO messages transmitted by $n_j$ between timestamps t1 and $t1+3\tau'$. Maximum number of HELLO messages of $n_j$ received by $n_k$ within the mentioned time interval is denoted as y and defined in (1). Significance of $\tau_{hello}(j)$ has been described earlier.

$$y = \lfloor 3\tau' / \tau_{hello}(j) \rfloor \qquad (1)$$

Let, the actual number of HELLO messages received by $n_k$ from $n_j$ between timestamps t1 and $t1+3\tau'$, be x. If x < y, then $n_i$ communicates with other downlink neighbours of $n_j$ at that time, to extract as many HELLO messages of $n_j$ as possible, within the mentioned time interval. Assuming that $x'$ number of such HELLO messages are collected and $n_j$ was found to be within radio-range of $n_i$ in $x''$ number of those messages, $n_i$ decides to blacklist $n_j$ network-wide, provided $x' = x'' = y$. Thereafter, each concerned honest downlink neighbour of $n_j$ is felicitated by $\alpha^2$ reputation to all other nodes in the network. On the other hand, if $x' < y$, reputation of $n_j$ to $n_i$, is modified according to rule bases of Adaptive Penalty Decider (APD). Accumulating all those HELLO messages if it is found that some downlink neighbour $n_l$ of $n_j$ tried to conceal at least one HELLO message of $n_j$ within the interval t1 and $t1+3\tau'$, then $n_l$ is also blacklisted network-wide by $n_i$. Otherwise, $n_l$ gains $\alpha^2$ increment in reputation to $n_i$ only, if allegation of link-breakage attack against $n_j$ is not confirmed. APD accepts the following input parameters and produces reputation-adjustment-quotient (RAQ) as output.

   i)   Comparative Reputation – Let $R_i^{min}$ and $R_i^{max}$ indicate minimum and maximum reputation respectively of all nodes who have communicated with $n_i$ so far, cooperative reputation $C_i(j)$ of a node $n_j$ to $n_i$, is defined in (2).

$$C_i(j) = (R_i(j) - R_i^{min})/( R_i^{max} - R_i^{min} +1) \quad (2)$$

Here $R_i(j)$ is the current reputation of $n_j$ to $n_i$. High comparative reputation $C_i(j)$ of a node $n_j$ increases its trustworthiness to its predecessor $n_i$. Please note that, $0 \leq C_i(j) < 2|\Phi|/(2|\Phi|+1)$.

ii) Expectation – Let $T_j(i)$ and $F_j(i)$ respectively denote amount of traffic node $n_j$ has generated to $n_i$ and total number of message packets $n_i$ has forwarded for $n_j$, so far. Expectation $E_j(i)$ of node $n_i$ from $n_j$ is defined in (3). As per this formulation, we can conclude that $0 \leq E_i(j) < 1$.

$$E_j(i) = F_j(i) / (T_j(i) + 1) \quad (3)$$

In spite of high values of $E_j(i)$, if $n_j$ does not cooperate with $n_i$, acceptability of $n_j$ to $n_i$ significantly decreases.

iii) Correctness – Correctness $Z_i(j)$ of the decision of punishment of $n_j$ by $n_i$, is expressed mathematically in (4). All symbols carry their usual meaning.

$$Z_i(j) = 1 – ( 1 - x'' / (x' + 1) ) ( 1 - x' / (y + 1) ) \quad (4)$$

From (4), it may be noted that $0 \leq Z_i(j) < 1$. Correctness more than 0.5 signifies that, $n_j$ was present within radio-range of $n_i$ in about more than half of the occasions after the last message packet was transmitted repeatedly (for 3 times) by node $n_i$ to $n_j$. This recommends strong penalty against $n_j$ in the form of its reduction in reputation in its predecessor.

iv) Path Length – Let p indicate the total number of nodes from source to destination in the associated communication path and q is the same before current silent router $n_j$, who is suspected to have issued the link breakage attack. Then, fractional length of path $P(j)$ traversed up to node $n_j$, appears in (5).

$$P(j) = q / p \quad (5)$$

Please note here, $1/H < P_i(j) < 1$, where H is the maximum hop count in the network. Low values of q signify that the destination is still far away and route-request packet generated by the predecessor of $n_j$ will have to traverse a substantially long way to reach the destination, compared to total length of the path. Hence, values of P closer to 0 claim severe punishment for $n_j$.

Divisions in crisp ranges of the above parameters along with associated fuzzy variables are shown in table I. Assuming that $f(\Phi)=2|\Phi|/(2|\Phi|+1)$, comparative reputation C is divided into four uniform ranges between 0 and $f(\Phi)$. Similarly, range division of P and E take place uniformly over the interval [1/H , 1] and [0 , 1], respectively. As far as range distribution of correctness Z is concerned, it is quite strict. At least 50% of this parameter is required to decrease reputation of the node suspected to issue link breakage attack, to its predecessor.

Table I : Crisp Range Division of C, E, Z and P Along With Fuzzy Variables

| Crisp Ranges of E, RAQ | Crisp Ranges of C | Crisp Ranges of Z | Crisp Ranges of P | Fuzzy Premise Variables |
|---|---|---|---|---|
| 0 – 0.25 | 0 - $f(\Phi)/4$ | 0 - 0.50 | 1/H – 1/4(1+3/H) | a |
| 0.25-0.50 | $f(\Phi)/4$ - $f(\Phi)/2$ | 0.50 – 0.65 | 1/4(1+3/H) – 1/2(1+1/H) | b |
| 0.50-0.75 | $f(\Phi)/2$ - $3f(\Phi)/4$ | 0.65 – 0.85 | 1/2(1+1/H) - 1/4(3+1/H) | c |
| 0.75-1.00 | $3f(\Phi)/4$ – 1.00 | 0.85 – 1.00 | 1/4(3+1/H) – 1.00 | d |

Fuzzy composition of C and E is shown in table II that produces temporary output ρ1. ρ1 is combined with Z in table III generating another intermediate output ρ2. Ultimate output RAQ is produced as per table IV as a function of ρ2 and P.

Table II : Fuzzy Composition of C and E producing ρ1

| C → <br> E ↓ | a | b | c | d |
|---|---|---|---|---|
| a | b | b | a | a |
| b | c | b | b | b |
| c | d | c | c | c |
| d | d | d | d | d |

Table III : Fuzzy Composition of Z and ρ1 producing ρ2

| Z → <br> ρ1 ↓ | a | b | c | d |
|---|---|---|---|---|
| a | a | b | b | c |
| b | a | b | c | c |
| c | a | b | c | d |
| d | b | c | d | d |

Table IV : Fuzzy Composition of P and ρ2 producing RAQ

| P → <br> ρ2 ↓ | a | b | c | d |
|---|---|---|---|---|
| a | b | a | a | a |
| b | c | b | b | b |
| c | d | c | c | c |
| d | d | d | d | d |

Updated reputation of $n_j$ to $n_i$ is given by (6).

$$R_i(j) = \begin{cases} R_i(j)\,\kappa(RAQ) & \text{if total number of occasions for which } n_j \text{ was suspected to issue link} \\ & \text{breakage attack against } n_i, \text{ is less than M} \\ -|\Phi| & \text{Otherwise} \end{cases} \quad (6)$$

Here $\kappa(RAQ) = ((RAQ)_{low} + (RAQ)_{high})/2$ and $RAQ \in \{a,b,c,d,e\}$. As far as crisp range division of RAQ is concerned, $a_{low} = 0$, $a_{high} = 0.25$, $b_{low} = 0.25$, $b_{high} = 0.50$, $c_{low} = 0.50$, $c_{high} = 0.75$ and $d_{low} = 0.75$, $d_{high} = 1.00$. M is the maximum number of times $n_j$ may be suspected to have raised link breakage attacks against $n_i$, before $R_i(j)$ reaches $-|\Phi|$.

*Proposition 1.* In a communication path … $n_i \rightarrow n_j \rightarrow n_k$ …, if $n_k$ is selfish, it surely provides $n_i$ correct information regarding the HELLO messages transmitted by $n_j$, during the time interval $t1$ and $t1+3\tau'$ (all symbols carry their usual meaning). On the other hand, if $n_k$ is malicious, the damage that it may cost to the entire network is bounded.

Case-1  $n_k$ is selfish

Let $\psi_1$ be the probability that $n_i$ has found a route to $n_k$ excluding $n_j$. Probability that $n_j$ is blacklisted, is given by $\psi_2$. $\vartheta^h(k)$ and $\vartheta^d(k)$ denote gain in reputation of $n_k$ through cooperation and selfishness, respectively. Mathematical expressions for these appear in (7) and (8). Please note here that if $n_j$ could be blacklisted, reputation of $n_k$ increases by $\alpha^2$ to all nodes in the network except itself and $n_j$. Based on the assumption that $\Phi$ is the set of all nodes in the network, resultant gain in reputation for $n_k$ in this situation, is $\alpha^2$ ($|\Phi|$-2). On the other hand, if $n_j$ could not be blacklisted, reputation of $n_k$ increases by $\alpha^2$ to $n_i$ only. If $n_i$ does not find any route to $n_k$ excluding $n_j$, $R_i(k)$ does not change. As far as $\vartheta^d(k)$ is concerned, since $n_k$ has decided to conceal at least one HELLO message transmitted by $n_j$ within time interval t1 and t1+3$\tau'$, network-wide blacklisting of $n_j$ becomes impossible. Hence, reputation of $n_k$ increases by $\alpha^2$ to $n_i$ only.

$\vartheta^h(k) = \psi_1\{\psi_2\alpha^2 (|\Phi|-2) + (1-\psi_2) \alpha^2\} + (1-\psi_1)0$
i.e. $\vartheta^h(k) = \psi_1\{\psi_2\alpha^2 (|\Phi|-2) + (1-\psi_2) \alpha^2\}$  (7)

$\vartheta^d(k) = \psi_1\alpha^2 + (1-\psi_1)0$
i.e. $\vartheta^d(k) = \psi_1\alpha^2$  (8)

Please note that, $\vartheta^h(k) - \vartheta^d(k) = \psi_1\psi_3\alpha^2 (|\Phi|-3)$ (surely, $|\Phi| > 3$)  (9)
From practical point of view, we can say that $|\Phi| > 3$, hence $(\vartheta^h(k) - \vartheta^d(k)) > 0$.
This ensures that a selfish node never breaks a communication link deliberately.

Case-2  $n_k$ is malicious

Objective of a malicious node is to cause as much harm as possible to other network elements till the consequences are bearable to itself. If $n_k$ conceals some HELLO messages of $n_j$ transmitted between timestamps t1 and t1+3$\tau'$, from $n_i$, intentional misbehaviour of $n_j$, if at all, will surely escape detection. Also $n_k$ undergoes the risk of getting isolated from the network. Even if malice of $n_k$ remains undetected, upper bound of damage it may cause to the network is given by HM$\sigma$($|\Phi|$-1) where H, M and $\Phi$ carry their usual meaning. Given that $D_{max}$ indicate the maximum of all radio-ranges in the network, $\sigma$ is the approximate energy required to transmit a message by 1-hop where the associated downlink neighbour is at a distance of $D_{max}$ from its predecessor. Hence, it is proved that if a node is malicious and tries to raise link breakage attack, the damage that it may cause to the network, is bounded.

### 3.3. Deliberately Delaying The Traffic

In a communication path … $n_i \rightarrow n_j \rightarrow n_k$ …, let t1 be the timestamp when $n_i$ relayed a message to $n_j$ and $n_j$ forwarded the same to $n_k$ at time t2. After $n_j$ receives a message and stores in the message queue, maximum possible waiting time in the queue is $(m_j-1)\tau$ where $\tau$ is the uniform time span required to process and forward a message and $m_j$ is size of the message queue of $n_j$. $n_i$ broadcasts network-wide allegation against $n_j$ for deliberately delaying the traffic provided $(t2-t1)>((m_j-1)\tau+3\tau')$; all symbols carry their usual meaning. Implicit acknowledgements or received copy of bicasted message is attached as proof along with the complaint. Otherwise reputation of $n_j$ to $n_i$ is adjusted by APD again. In this case it accepts all parameters except path length P and correctness Z is redefined as (10).

$Z_i(j) = (t2-t1)/((m_j-1)\tau+3\tau')$  (10)

Here also Z lies between 0 and 1. Values closer to 1 emphasize punishment of $n_j$. As far as deliberately delaying is concerned, Z follows uniform range distribution like E and RAQ. Output is produced from table III i.e. $\rho2$ = RAQ. For a communication path r, assuming that $\lambda(r)$ be the total number of routers in it, time interval $\Gamma(r)$ for which source node waits for arrival of acknowledgement from destination is given by expression (11).

$\Gamma(r) = (\lambda(r)+1) \tau + (\tau+3\tau')\lambda(r) + \sum_{n_l \in \lambda(r)} (m_l-2) \tau$  (11)

Here $(\lambda(r)+1)\tau$ is the lower bound of waiting time before arrival of acknowledgement from destination, while rest of the expression (11) is the upper bound. If acknowledgement does not arrive within interval $\Gamma(r)$, each node voluntarily returns received copy of the message bicasted by itself, to their respective predecessors. Same proactive procedure is followed by the nodes for which link with the predecessor is about to be broken before time interval $\Gamma(r)$. If there exists any node $n_i$ who did not receive the said proof of cooperation from its successor $n_j$, tries to establish a communication path to successor of $n_j$ excluding $n_j$ itself, to inquire whether $n_j$ forwarded the message to it. If it is found that $n_j$ has forwarded the message properly, $R_i(j)$ is incremented by $\alpha$, otherwise $R_i(j)$ is minimized. If no such alternative path could be established from $n_i$ to successor of $n_j$ and it happened for M number of occasions so far, then also $R_i(j)$ is minimized.

### 3.4. Injecting Enormous Traffic

RACS allows a maximum of $\eta$ route-requests per second. In our simulation, we have set $\eta = 5$. If any downlink neighbour $n_j$ of $n_i$ receives more than $\eta$ number of route-requests from $n_i$ in one second (may be in different destinations), $n_j$ broadcasts an allegation against $n_i$ with last $\eta+1$ route-requests as proof. Timestamps embedded in those messages establish the fact that they were broadcasted within time stamp of one second. Thus $n_i$ is isolated from the network.

### 3.5. Collusion Attack

Collusion is the act of cooperating with another node's misbehaviour. According to RACS, selfish and malicious nodes deal with collusion requests differently. As far as RACS is concerned, possibility of collusion lies in the following situations.
   i)   In a communication path … $n_i \rightarrow n_j \rightarrow n_k$ …, $n_j$ raises a link breakage attack against $n_i$ and asks $n_k$ to collude with it by concealing some HELLO messages of $n_j$ transmitted between the time interval t1 and t1+3$\tau'$ (t1 and $\tau'$ carry their meaning as in section 3.2).
   ii)  Neighbours of a node $n_i$ may collectively decide not to forward its messages and thereby isolate it from the network.

Situation i has already been discussed in section 3.2. Proposition 2 below proves that, irrespective of whether a node is selfish or malicious, gain through honesty is higher than the gain through collusion.

*Proposition 2. Irrespective of whether a node is selfish or malicious, it is a better option to raise network-wide allegation against network element issuing malicious collusion requests.*

*Proof :* Assume that $n_j$ and $n_k$ are two network elements s.t. $n_j, n_k \in N(i)$. $n_j$ issues a collusion request to $n_k$ for conspiring against $n_i$.

Case-1  $n_k$ is selfish

Let $\vartheta^h(k)$ and $\vartheta^d(k)$ denote the gain of $n_k$ possible through honesty and colluding with $n_j$, respectively. Assuming that $\Phi$ is the set of all nodes in the network, mathematical expressions for $\vartheta^h(k)$ and $\vartheta^d(k)$ are computed in (12) and (13). If $n_k$ decides to behave honestly, it will raise network wide allegation against $n_j$ with received copy of collusion request from $n_j$ against $n_i$ as supporting digital document. This will blacklist $n_j$ and fetch a reputation of $\alpha^2$ to all other network elements except $n_j$ and itself i.e. resultant gain in reputation amounts to $\alpha^2$ ($|\Phi|$-2). On the other hand, if $n_k$ colludes with $n_j$ in its conspiracy against $n_i$, it gains reputation $\alpha^2$ to $n_j$ only.

$$\vartheta^h(k) = \alpha^2 (|\Phi|-2) \qquad (12)$$

$$\vartheta^d(k) = \alpha^2 \qquad (13)$$
$$\vartheta^h(k) - \vartheta^d(k) = \alpha^2 (|\Phi|-3)$$

Again, from practical point of view, we can say that $|\Phi| > 3$, hence $(\vartheta^h(k) - \vartheta^d(k)) > 0$.

Case-2  $n_k$ is malicious

If $n_k$ decides to collude with $n_j$, the situation is equivalent to issuing link breakage attack against $n_i$ or some 2-hop uplink neighbour of $n_i$. We have already proved in proposition 1 that damage caused by $n_j$ to the network is bounded, for both the above-mentioned situations. Moreover, the risk of getting caught red-handed is there. On the other hand, if $n_k$ raises network wide allegation against $n_j$, it can surely blacklist $n_j$ for ever and the effect is much more severe than that of link breakage attack. Moreover, it fetches benefit for $n_k$ in terms of reputation, which is mathematically expressed in (12).

### 3.6. Slander attack

If a node $n_i$ raises allegation against another node $n_j$ without proof, then it is not entertained and $n_i$ is blacklisted throughout the network.

### 3.7. Masking attack

Whenever a node generates a message, its unique node identifier automatically gets embedded in it, which cannot be changed. As per this constraint of RACS, a node cannot pretend to be another one.

## 4. ALGORITHM

The communication algorithm in RACS is presented below in detail in procedure communication where $n_i$ is sending a message to $n_j$ for forwarding. It illustrates the technique to deal with link breakage attack and deliberately delaying traffic attack. On the other hand, whenever some route-request message arrives at $n_j$ from $n_i$, $n_j$ behaves according to procedure route_req to detect injecting enormous traffic attack. Pseudo code of 'route_req' and the procedures called from within the procedure 'communication' are presented in the appendix A.

```
Procedure communication (node n_i, n_j, n_s, n_d)
begin
 /* this procedure illustrates communication from n_i to n_j, from the perspective of n_i after a path is
established between them. n_s and n_d are source and destination node respectively */
 flag = 0
 /* flag is a Boolean variable which is assigned 0 initially. It will be set to 1 if a reply is received from n_j
after at most 3 retransmissions */
  if (n_j ∉ n_i.comm_history) then
  begin
 /* n_i is communicating with n_j for the first time*/
   reputation_i[j] = 0;
   link_break_i[j] = 0;
   deliberate_delay_i[j] = 0;
   blacklist_i[j] = 0;
  end
t1 = sys_time();
/* sys_time is a function which returns current timestamp of the system*/
 send_msg(n_i, n_i, n_j, t1);
/* the message is transmitted by n_i for the first time at timestamp t1. It is bicasted to both n_i and n_j */
 reply1 = receive_msg(n_i, n_i, t1);
```

```
/* reply1 indicates copy of the bicasted message transmitted and received by n_i */
 reply = null;
  for p = 1 to 3
   begin
     reply = receive_ack (n_i, n_j, sys_time());
     /* if any acknowledgement arrives from n_j to n_i, then it is stored in a variable named
        reply. If no reply comes then the message will be resent and this may continue at
        most three times */
      if (reply is not null) then
       begin
         flag = 1;
         break;
       end
      else
       begin
         t_last = sys_time();
         send_msg(n_i, n_j, sys_time());
         /* during retransmissions, the message is unicasted to next hop router */
       end
    end
    t2 = sys_time();
    z = ⌈ (t2 – t1) / τ_hello(j) ⌉ ;
    /* z is the estimated number of hello messages of n_j between timestamps t1 and t3 */
    if ( flag = 0) then
   begin
    /* n_j has kept mum to message forwarding requests sent by n_i */
    black_count = 0;
    /* black_count will store number of hello messages which support blacklisting of n_i
       throughout the network */
    all_reply = null;
    /* all_reply will consists of responses of all members of N(j), where N(j) denotes
       current set of downlink neighbours of n_j */
    xarray = yarray = tarray = null;
    coord_count = 0;
    /* tarray is the array which will consist of unique timestamps among responses of all
       downlink neighbours of n_j. xarray and yarray will store corresponding positions of n_j
       in terms of x and y coordinates. Cardinality of all these arrays will be indicated as
       coord_count */
     for each node n_k ∈ N(j)
      begin
        hello_count[k] = 0;
        /* hello_count[k] is supposed to store the number of relevant hello messages of n_j
           forwarded to n_i by n_k */
        send_helloenq(n_i, n_j, n_k, t1,t2,sys_time());
        /* n_i sends an enquiry to all n_k ∈ N(j) regarding hello messages of n_j between
           timestamps t1 and t2 */
        reply_dlnebr = recv_reply(n_i, n_j, n_k, sys_time());
        /* reply_dlnebr is a variable is a variable which stores reply of n_k in response to
           enquiry of n_i regarding relevant hello messages of n_j, as mentioned in the
           message send_helloenq */
        all_reply = all_reply ∪ reply_dlnebr;
        /* response of n_k is appended to the variable all_reply */
        for l = 1 to z
         begin
           if (exhausted(reply_dlnebr) = 1)
            break;
           /* exhausted is a function which accepts an array as argument and returns 1 if end
              of array is reached, otherwise it returns 0 */
```

```
            hello_count[k] = hello_count[k] + 1;
         /* more hello messages of n_j are left in response of n_k to enquiry of n_i */
            info[k][l] = extract_nexthello (reply_dlnebr, l);
         /* extract_nexthello is a function which accepts response of n_k and current value of
            loop counter variable l, as argument. Its purpose is t extract l-th hello message of
            n_k from reply_dlnebr and store it in info[k][l] */
      /* each element of info consists of three attributes:
            i)      suspected_x : x coordinate of the node suspected to have raised link breakage attack
                    (here it is n_j)
            ii)     suspected_y : y coordinate of the node suspected to have raised link breakage attack
                    (here it is n_j)
            iii)    tmpstamp : timestamp of info[k][l]
            found = 0;
         /* found is set to 1 if info[k][l] has already been processed */
            for e =1 to coord_count
              begin
                if info[k][l].tmpstamp = tarray[coord_count] then
                 begin
                    /* this hello message has already been processed */
                    found = 1;
                    break;
                  end
                end
            if (found = 1) then continue;
            xarray[coord_count] = info[k][l].suspected_x;
            yarray[coord_count] = info[k][l].suspected_y;
            tarray[coord_count] = info[k][l].tmpstamp;
            coord_count = coord_count + 1;
          /* individual attributes of info[k][l] are appended to xarray, yarray and tarray;
             hence coord_count is also incremented */
            d = √(n_i.x- info[k][l].suspected_x)² +(n_i.y- info[k][l].suspected_y)²;
          /* d is the distance between n_i and n_j at time info[k][l].tmpstamp */
            if (d < radio_range(i)) then
          /* it is concluded here that n_j was within radio-range of n_i at time
             info[k][l].tmpstamp */
            black_count = black_count + 1;
           end
         end
      while not (exhausted(all_reply)) do
      /* the loop continues till response of all downlink neighbours of n_j has been processed
         by n_i */
        begin
         s = -1;
         /* s is a variable that is supposed to store number of honest downlink neighbours of
            n_j */
   for each node n_k ∈ N(j)
    begin
      hello_cnt[k] = 0;
      /* hello_cnt[k] is supposed to store the number of relevant hello messages of n_j
         that should have been forwarded to n_i by n_k */
      tot_record[k] = -1;
      tmvar = search_time (all_reply, n_k,0);
      /* search_time (all_reply, n_k,0) extracts timestamp of first (0-th) hello message of n_j
         from response of n_k. It is stored in variable tmvar */
         while tmvar ≠ 0 do
           begin
             tot_record[k] = tot_record[k] + 1;
             tmp[k][ tot_record[k]] = tmvar;
```

```
            xcoord[k][ tot_record[k]] = search_posx (all_reply, n_k, tmvar);
            /* search_posx is a function which returns x coordinate of n_k at time tmvar */
            ycoord[k][ tot_record[k]] = search_posy (all_reply, n_k, tmvar);
            /* search_posy is a function which returns y coordinate of n_k at time tmvar */
            xco_ord = search_posx (all_reply, n_j, tmvar);
            yco_ord = search_posy (all_reply, n_j, tmvar);
            /* xco_ord and yco_ord store x and y coordinate respectively, of n_j at time
               tmvar */
            dist1 = (xco_ord- xcoord[k][ tot_record[k]])²;
            dist2 = (yco_ord- ycoord[k][ tot_record[k]])²;
            dist = √(dist1+dist2);
           /* dist indicates distance between n_j and n_k at time tmvar */
            if (dist <radio_range(j)) then
           /* here it is concluded that n_k was neighbour of n_j at time tmvar */
            hello_cnt[k] = hello_cnt[k] + 1;
            tmvar = search_time (all_reply, n_k,tmvar);
            /* new value of tmvar is the next timestamp extracted by search_time function
                from all_reply, after value of tmvar in previous pass through the loop */
          end
        end
      if (hello_cnt[k] = hello_count[k]) then
      /* n_k did not conceal any relevant hello message of n_j from n_i */
      begin
        s = s + 1;
        arr [s] = k;
        /* n_k is appended in the list of honest downlink neighbours of n_j*/
       end
     end
    end
  sum_hello_found = 0;
  for each node n_k ∈ N(j)
    sum_hello_found = sum_hello_found + hello_count[k];
  /* sum_hello_found is total number unique hello messages received from all members
     of  N(j) */
  if ((black_count = z) or (black_count = z - 1)) then
    begin
     for each node n_m ∈ Φ
      /* Φ is the set of all nodes in the network */
         send_allegation_link (n_i, n_j, n_m, all_reply,arr,s,sys_time());
        /* n_i alleges n_j of raising link breakage attack while all_reply works as the
           supporting digital document. After receiving this allegation, n_m executes a
           procedure termed as receive_allegation_link. */
     end
   else
  /* n_j could not be blacklisted. So, its reputation to n_i would be set to -Φ if this kind of situation occurred
  for M number of times as far as the communication from n_i to n_j is concerned. Otherwise, the fuzzy
  controller named adaptive penalty decider is invoked for link breakage scenarios through a function
  APD1 */
   begin
    link_break_i[j] =  link_break_i[j]  + 1;
    if link_break_i[j] > M then
       reputation_i[j] = -Φ;
    else
       reputation_i[j] = APD1 (black_count, sum_hello_found, n_j, n_s, n_d);
   end
  end
  else
   begin
```

```
 /* reply is available from n_j */
  t3 = retrive_timestamp (reply, n_j);
  m_j = retrive_sizeofqueue (reply, n_j);
 /* t3 is the timestamp when n_j forwarded the message from n_i to its successor in the communication path
*/
/* m_j is the size of message queue of n_j */
 if ((t3-t_last)>( m_j -1) τ) then
  begin
   /* n_j has deliberately introduced delay in the traffic */
    for each node n_m ∈ Φ
     /* Φ is the set of all nodes in the network */
       send_allegation (n_i, n_m, n_j, reply, reply1,sys_time());
   /* After receiving this allegation, n_m executes a procedure termed as
       receive_allegation */
       blacklist_m[j] = 1;
     /* n_j is blacklisted to n_m */
      reputation _m[j] = -Φ;
      /* reputation of n_j is minimized to n_m */
      reputation _m[i] = Φ;
     end
    end
  else
  /* n_j could not be blacklisted. So, its reputation to n_i would be set to -Φ if this kind of situation occurred
for M number of times as far as the communication from n_i to n_j is concerned. Otherwise, the fuzzy
controller named adaptive penalty decider is invoked for link breakage scenarios through a function
APD2 */
  begin
   deliberate_delay_i[j] =  deliberate_delay_i[j]  + 1;
   if deliberate_delay_i[j] > M then
      reputation_i[j] = -Φ;
   else
     reputation_i[j] = APD2 (t3, t_last, n_j, reply, reply1);
    end
  end
end
```

## 5. SIMULATION RESULT

In our simulations, nodes are randomly deployed inside a rectangular space. Each node moves according to a random-waypoint mobility model. In this model, a node starts at a random position, waits for a random duration called the pause time, then randomly chooses a new location and moves to that location with random velocity ranging between 0 and maximum allowable velocity in the network. When it arrives at the new location it waits for a random pause time and repeats the process. The physical layer assumes a transmission range model fixed for individual nodes although transmission ranges of different nodes are different. One node can successfully communicate in single hop with another node provided the second node is within radio-range of the first one. Important simulation parameters are listed in table V. We have used ns2 [15] simulator.

Each selfish node acts as a service provider, which randomly picks another node as receiver. Packets are scheduled to be generated according to a Poisson process. Total number of malicious nodes varies from 0 to 500 in different simulation runs. Anyone of them randomly picks a strategy of causing damage to network.

Attributes $attr_i$ of a node $n_i$ are as follows:
i) node_id : i
ii) latitude : $lat_i$
iii) longitude : $long_i$
iv) radio-range : $rad_i$
v) velocity : $v_i$
vi) interval between
consecutive hello messages : $\tau_{hello}(i)$

Table V:   Important Simulation Parameters

| | |
|---|---|
| Total number of nodes | 500 - 5000 |
| No. of malicious nodes | 0 – 500 |
| Dimension of space | 2000 m × 1000 m |
| Maximum radio-range | 250 m |
| Network-wide maximum velocity | 50 m /s |
| Network-wide minimum interval between consecutive HELLO messages | 6 seconds |
| Network-wide maximum interval between consecutive HELLO messages | 10 seconds |
| Interval between consecutive retransmissions of a message | 10 seconds |
| Maximum number of route_requests per second | 5 |
| Data and control packet size | 512 bytes |
| Link bandwidth | 1 Mbps |
| Simulation Time | 3600 seconds |
| Number of simulation runs | 6 |

As per the simulation scenario presented in table V, size of $attr_i$ is ($\log_2 5000 + \log_2 2000 + \log_2 1000 + \log_2 250 + \log_2 50 + \log_2 30$) bits i.e. 53 bits. $\tilde{N}(i)$ consists of first seven elements of $N(i)$ after they are arranged in decreasing order of proximity with $n_i$. Assuming that $U(i)$ indicates current set of uplink neighbours of $n_i$, $\tilde{U}(i)$ consists of first four elements of $U(i)$ after they are arranged in decreasing order of proximity with $n_i$. In all the message formats mentioned below, first 4 bits are message code which range from 0 to 14.

   Based on selfish nodes' forwarding decision, three systems have been implemented in the simulations: the proposed RACS system, CONFIDANT [4] and Attack-resistant Cooperation Stimulation [7]. All these mechanisms deal with both selfish and malicious nodes. Relevant performance metrics used here are described below. Among them, the first one is reputation efficiency of selfish nodes, which is the total profit gained by all selfish nodes (in terms of reputation for RACS and CONFIDANT [4] or energy for Attack-resistant Cooperation Stimulation [7]) divided by the total number of selfish nodes.  It is followed by average damage received per selfish node, which is the total damage (in terms of percentage of reputation for RACS and CONFIDANT or energy for Attack-resistant Cooperation Stimulation) received by selfish nodes divided by the total number of selfish nodes. The next performance metric is average damage received per malicious node, similar to the measure of Average damage received per selfish node. Last performance metric is correctness of detection of malicious nodes which is ((1-mal_det / act_mal ) × 100). mal_det signifies the total number of malicious nodes detected and act_mal is the actual number of malicious nodes participating in communication sessions of simulation.

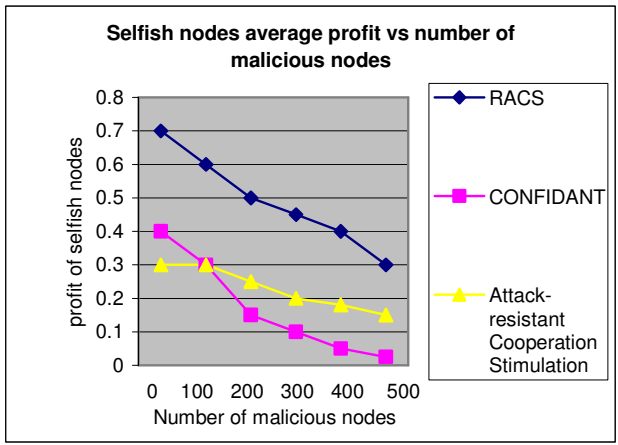

Fig 1. Graphical demonstration of selfish nodes average reputation efficiency vs number of malicious nodes

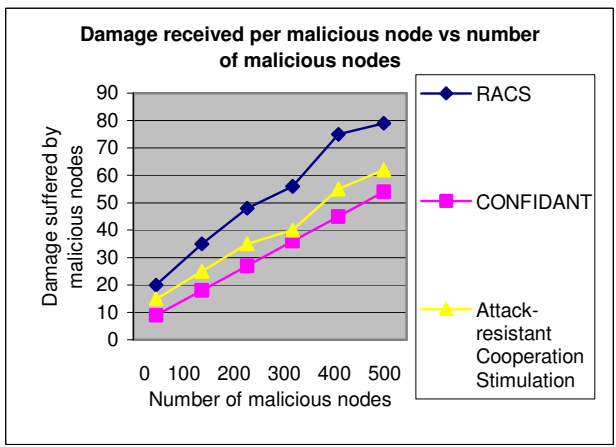

Fig 2. Graphical demonstration of damage received per malicious node vs number of malicious nodes

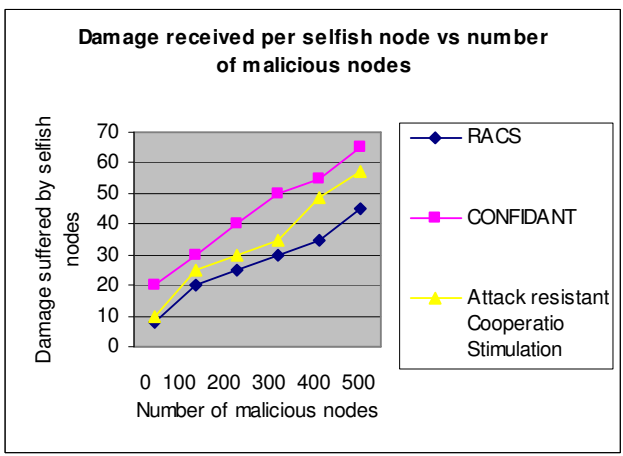

Fig 3. Graphical demonstration of damage received per selfish node vs number of malicious nodes

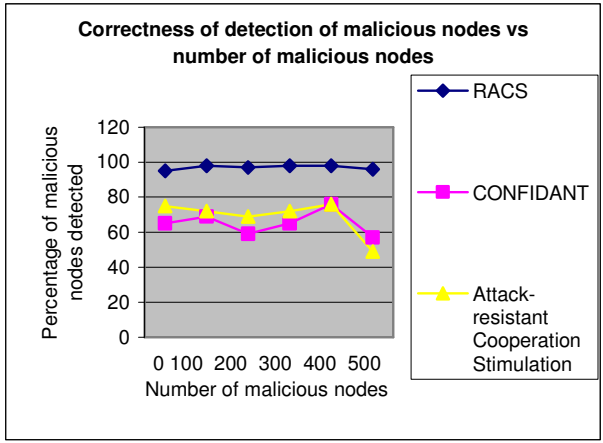

Fig 4. Graphical demonstration of correctness of detection of malicious nodes vs number of malicious nodes

Table VI : Improvement of RACS over CONFIDANT and Attack-resistant Cooperation Stimulation

| Performance metric | RACS over CONFIDANT | RACS over Attack-Resistant Cooperation Stimulation |
|---|---|---|
| Selfish nodes average profit vs number of malicious nodes | 33.32% | 31.45% |
| Damage received per malicious node vs number of malicious nodes | 37.19% | 32.1% |
| Damage received per selfish node vs number of malicious nodes | 38.67% | 27.32% |
| Correctness of detection of malicious nodes vs number of malicious nodes | 29.28% | 22.34% |

## 6. CONCLUSION

RACS is a reputation-based, completely distributed method of detecting and preventing selfish and malicious activities in ad hoc networks. In contrast to the earlier schemes, it employs nodes themselves to monitor activities of other nodes; need for intervention of any centralized body has been intelligently eliminated. The communication procedure has been described in details along with the structure of various categories of control and data packets. RACS discriminates between intentional and unintentional selfish and malicious actions. Nodes that detect malicious/selfish activities of others are recognized network-wide, whereas the culprits have to suffer depending upon their degree of non-cooperating and harmful attitude. Simulation results present in section 5, strongly establish the effectiveness of our proposed RACS system.

# APPENDIX A:

Pseudo code for the procedures receive_allegation and receive_allegation_link are presented below.

Procedure receive_allegation (node $n_i$, $n_j$, $n_m$, string reply, reply1, integer t)
 begin
  if(check(reply, reply1, $n_i$, $n_j$)=1) then
  /* check is a function that investigates guilt of $n_j$ according to the digital proof
     all_reply. It returns 1 if $n_j$ is found guilty. Otherwise it returns 0. */
   begin
     blacklist$_m$[j] = 1;
      /* $n_j$ is blacklisted to $n_m$ */

```
         reputation _m[j] = -Φ;
          /* reputation of n_j is minimized to n_m */
         reputation _m[i] = Φ;
          /* reputation of n_i is maximized to n_m because n_i has detected misbehaviour of n_j */
          /* nodes n_arr[s] supported n_i in its procedure of detecting misbehaviour of n_j .So, their
         reputation is also maximized to n_m */
        end
     else
      begin
         blacklist_m[i] = 1;
          /* n_i is blacklisted to n_m */
         reputation _m[i] = -Φ;
          /* reputation of n_i is minimized to n_m */
        end
       end

Procedure receive_allegation_link (node n_i, n_j, n_m, string all_reply, integer array arr, integer s, integer t)
 begin
   if(check(all_reply,null, n_i, n_j)=1) then
    /* check s a function that investigates guilt of n_j according to the digital proof all_reply.
       It returns 1 if n_j is found guilty. Otherwise it returns 0. */
     begin
         blacklist_m[j] = 1;
          /* n_j is blacklisted to n_m */
         reputation _m[j] = -Φ;
          /* reputation of n_j is minimized to n_m */
         reputation _m[i] = Φ;
          /* reputation of n_i is maximized to n_m because n_i has detected misbehaviour of n_j */
          /* nodes n_arr[s] supported n_i in its procedure of detecting misbehaviour of n_j .So, their
            reputation is also maximized to n_m */
             for t = 0 to s-1
                 reputation _m[arr[t]] = Φ;
        end
     else
      begin
         blacklist_m[i] = 1;
          /* n_i is blacklisted to n_m */
         reputation _m[i] = -Φ;
          /* reputation of n_i is minimized to n_m */
        end
       end

Procedure route_req (node n_j, n_i, n_s, n_d, integer tm)
begin
/* this procedure illustrates how n_j deals with route request message from n_i transmitted at time tm */
 msg = store_rreq(node n_j, n_i, n_s, n_d, tm);
 /* the entire route request is stored in a variable termed msg by store_rreq function*/
 if ((n_i ∉ n_j.rreq_history)  or  (tm-time[i][j] > 1)) then
  begin
   /* n_i has sent a route-request to n_j for the first time or after at least one second from the earlier route-
request */
    rreq[i][j] = 0;
 /* (rreq[i][j]+1) is the number of route requests of n_i to n_j after at least one second from the earlier route-
request */
    time[i][j] = tm;
```

```
  setoff_req = null;
/* time[i][j] is timestamp of last route request message of n_i to n_j */
 end
else
 begin
  if (rreq[i][j] < 5)
   begin
    rreq[i][j] = rreq[i][j] + 1;
    append (setof_rreq, msg);
   /* setof_rreq is the set of route requests of n_i to n_j after at least one second from the
      earlier route-request */
    self_or_mal = random(1);
    if (self_or_mal = 1)
     /* n_j will forward the route request message because it has decided to cooperate
        with n_i */
        forward (n_j, n_s, n_d, sys_time());
   end
   else
    begin
      for each node n_m ∈ Φ
     /* Φ is the set of all nodes in the network */
        send_allegation (n_i, n_j , n_m, setof_req, sys_time());
     /* After receiving this allegation, n_m executes a procedure termed as
        receive_allegation */
     end
    end
end
```

**Authors**

Anuradha Banerjee is working as a lecturer in kalyani Govt Engg College in the department of Computer Applications since 2003. She completed her B.E. from Bengal Engineering College, Sibpur. She has 22 publications to her credit. She is doing Ph.d. on ad hoc networks.

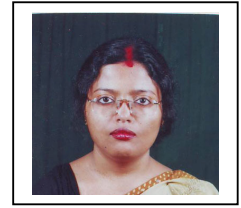

Dr. Paramartha Dutta is currently working as a Professor in Visva-Bharati University, Santiniketan. He has completed his M.Tech. from Indian Statistical Institute, Calcutta in 1993 and Ph. D. in Computer Science and Engineering from Bengal Engineering And Science University, Sibpur, West Bengal, India in the year 2005. He has 63 publications in to his credit.

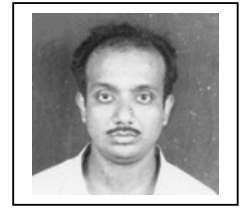